\documentclass[reprint,amsmath,amssymb,aps,noeprint,floatfix, nolinenumbers,longbibliography]{revtex4-2}

\usepackage[english]{babel}
\usepackage[utf8]{inputenc}
\usepackage[colorinlistoftodos, color=green!40, prependcaption]{todonotes}
\usepackage{float}
\usepackage{graphicx}
\usepackage{yhmath}
\usepackage{subfigure}
\usepackage{mathdots}
\usepackage{MnSymbol}
\usepackage{dsfont}
\usepackage{soul}
\usepackage{ulem}
\usepackage{hyperref}
\usepackage[sort&compress]{natbib}

\hypersetup{
    unicode=true, 
    plainpages=false,
    colorlinks=true,
    linkcolor=blue,
    citecolor=blue,
    filecolor=blue,
    urlcolor=blue
}

\usepackage{mathtools}

\DeclarePairedDelimiter\ket{\lvert}{\rangle}
\DeclarePairedDelimiterX\braket[2]{\langle}{\rangle}{#1\,\delimsize\vert\,\mathopen{}#2}
\newcommand{\nb}[1]{\textcolor{black}{#1}}

\newcommand{\op}[1]{\hat{#1}}
\def\dd{\mbox{d}}

\begin{document}

\title{Ising Hamiltonian Minimization: Gain-Based Computing with Manifold Reduction of Soft-Spins vs Quantum Annealing}

\author{James~S.~Cummins${}^1$, Hayder Salman${}^{2, 3}$ and Natalia~G.~Berloff${}^{1,}$}
\email[correspondence address: ]{N.G.Berloff@damtp.cam.ac.uk}
\affiliation{${}^1$Department of Applied Mathematics and Theoretical Physics, University of Cambridge, Wilberforce Road, Cambridge CB3 0WA, United Kingdom\\
${}^2$School of Engineering, Mathematics, and Physics, University of East Anglia, Norwich Research Park, Norwich, NR4 7TJ, UK,\\
${}^3$Centre for Photonics and Quantum Science, University of East Anglia, Norwich Research Park, Norwich, NR4 7TJ, UK}

\date{\today}

\begin{abstract}
We investigate the minimization of Ising Hamiltonians, comparing the performance of gain-based computing paradigms based on the dynamics of semi-classical soft-spin models with quantum annealing. We systematically analyze how the energy landscape for the circulant couplings of a M\"{o}bius graph evolves with increased annealing parameters. Our findings indicate that these semi-classical models face challenges due to a widening dimensionality landscape. To counteract this issue, we introduce the `manifold reduction' method, which restricts the soft-spin amplitudes to a defined phase space region. Concurrently, quantum annealing demonstrates a natural capability to navigate the Ising Hamiltonian's energy landscape due to its operation within the comprehensive Hilbert space. Our study indicates that physics-inspired or physics-enhanced optimizers will likely benefit from combining classical and quantum annealing techniques.
\end{abstract}

\maketitle

\section{Introduction}

Pursuing enhanced computing speed and power efficiency has led to exploring alternatives to traditional electronic systems in solving complex tasks. Optical Neural Networks (ONNs) promise unprecedented parallelism, potentially superior speeds, and reduced power consumption. ONNs encode neural weights as phase shifts or changes in light intensity, with activation functions instantiated via nonlinear optical materials or components, or via a strong hybridization to matter excitations \cite{kasprzak2006bose}. They offer the potential to operate in the terahertz range, vastly surpassing the gigahertz frequencies of conventional electronic systems that can be exploited in machine learning and combinatorial optimization. The common feature of ONNs is to utilize a network of optical oscillators dynamically described by a coupled system of soft-spin models on complex-valued fields $\psi_i = r_i \exp[i \theta_i]$ that have amplitude $r_i$ (referred to as `soft mode') and phase $\theta_i$ (discrete e.g. $\theta_i \in \{0, \pi\}$ or continuous `spin') degrees of freedom. Each spin in the network can be associated with the quadrature of the optical complex-valued fields, thereby in the classical limit reducing the system to a model of real soft-spins given by $r_i \cos \theta_i$ which we analyze hereafter.
 
Optical parametric oscillator based coherent Ising machines (CIMs) \cite{yamamoto2017coherent, inagaki2016coherent, mcmahon2016fully,Honjo2021}, lasers \cite{pal2020rapid, babaeian2019single, parto2020realizing}, spatial light modulators (SLMs) \cite{pierangeli2019large}, lattices of polariton \cite{berloff2017realizing, kalinin2020polaritonic} and photon condensates \cite{vretenar2021controllable}, Microsoft's analog iterative machine \cite{kalinin2023analog}, and Toshiba's simulated bifurcation machine \cite{goto2021high} can all minimize the classical hard-spin Ising Hamiltonians $H_{\rm I} = -\sum_{i,j} J_{ij} s_i s_j$ with $s_i = \pm 1$ for a coupling matrix $\mathbf J$, and other spin Hamiltonians (e.g. XY Hamiltonians $H_{\rm XY} = -\sum_{i,j} J_{ij} \mathbf{s}_i \cdot \mathbf{s}_j$ with $\mathbf{s}_i = (\cos \varphi_i, \sin \varphi_i)$) using soft-spin bifurcation dynamics via the Aharonov-Hopf bifurcation \cite{syed2022bifurcation}. This principle of operation leads to an exciting new paradigm known as ``gain-based computing''. The concept behind gain-based computing is that computational problems can be encoded in the gain and loss rates of driven-dissipative systems, which are then driven through a symmetry-breaking transition (bifurcation), selecting a mode that minimizes losses. Such soft-spin models exploit enhanced dimensionality, marked by small energy barriers during amplitude bifurcation, but also complicate the energy landscape with numerous local minima. In parallel to these methods, quantum annealing is another approach to minimize the hard-spin Ising Hamiltonian. Despite numerous studies contrasting classical and quantum methods, the limitations of currently available hardware and the limitations of simulating quantum systems classically have led to contrasting conclusions as to whether a quantum advantage can potentially be realized using quantum annealing \cite{Liu2015, King2022, Heim2015, King2023, Tameem2018, Yan2022, Albash2018, Muthukrishnan2016} and in particular, how quantum annealers such as D-Wave perform in comparison with CIM \cite{Hamerly2019}. In the latter, the connectivity of the coupling matrix was assumed to be a key factor in performance differences between these machines \cite{Hamerly2019}.

An all-optical scalable ONN was recently proposed for cyclic graphs; SLMs are used to discretize the optical field where each pixel defines a different pulse amplitude \cite{strinati2021optical}. The SLM with $M_{x} \times M_{y}$ pixels is set up with a transmission function $\tilde{J}_{k}$ which multiplies the Fourier transform of the amplitudes at each round trip. The SLM couples the fields with coupling matrix $J_{ij} \equiv \tilde{J}_{j-i + 1}$, which corresponds to a circulant graph. An alternative setup allows for any general coupling matrix. However, there is an $N = M_{x}$ limit to the number of pulses. Circulant graphs such as M\"{o}bius ladders therefore lend themselves well to optical solvers, where $N = M_{x} \times M_{y} \sim 10^{6}$ spins can be defined.

The couplings are often geometrical in polariton condensates, photon condensates, and laser cavities (e.g.\ the sign and amplitude of the coupling strength depend on the distance between condensates and outflow wavenumber \cite{ohadi2016nontrivial}). The condensates arranged in a circle interact with the nearest neighbors, but the interactions beyond this decay exponentially. Previously, various ways of establishing long-range interactions in polariton-based XY-Ising machines were discussed. An easier way to achieve the couplings between remote sites is to use digital micro-mirror devices (DMDs) to direct the light across the ring. DMDs were shown to perform complex (amplitude and phase) modulation. By splitting the complex field into real and imaginary parts and using the time modulation scheme of the DMD, a complex signal could be synthesized \cite{ayoub2021high}. Reflecting the entire ring of condensates on itself with a radial displacement implements a 3-regular cyclic graph. Cyclic graphs are known to be computationally intractable for classical computers for sampling probability distributions of quantum walks \cite{qiang2016efficient}.

Using ONNs for optimization has shown promise, yet key questions remain: `what are suitable benchmarks for optical machines, how to guide annealing to aid optimization, what are the ONN energy landscape dynamics during annealing to ensure the optimal state is achieved, and what are the distinguishing features between quantum and classical annealing?'. Answers often rely on the coupling matrix $\mathbf{J}$. An instructive problem encoded in $\mathbf{J}$ should be technologically feasible, have controllable couplings, possess non-trivial structures resistant to simple local perturbations, and be mathematically tractable. Moreover, it is better to have deterministic rather than random couplings to avoid issues of statistical convergence \cite{Marsh2023}. 

Here, we analyse and contrast gain-based computing for soft-spin Ising models (SSIM) with quantum annealing for circulant coupling matrices, which allow complete control of frustration, energy gaps, and the structure of critical points. Furthermore, the potential to realize them in future optical systems \cite{strinati2021optical} make them more suitable to consider over previously reported benchmarks \cite{hamze2020wishart, roberts2020noise, mandra2023generating}. A highlighted challenge for SSIM annealing lies in the opposing relationship between local and global minima when mapping the Ising Hamiltonian to the energy of the soft-spin system \cite{Marsh2023}. Notably, we demonstrate that quantum annealing within the whole Hilbert space of the hard-spin system navigates this challenge. Additionally, we suggest that `manifold reduction', aligning amplitudes to the mean, is needed to augment the likelihood of SSIM finding the global minima.

\nb{ONNs based on laser operation leverage quantum-inspired principles such as coherence, interference, and parallelism. They are dissipative systems that tend to minimize losses on their route to coherence.} The losses can be written as an `energy' (`cost') function to be minimized. For instance, in the classical limit, CIM's energy landscape to be minimized is
\begin{equation}
 E = \frac{C}{4} \sum_{i = 1}^{N} \left( p(t) - x_i^2 \right)^2 - \frac{1}{2} \sum_{i, j = 1}^{N} J_{ij} x_i x_j, \label{E_Ising}  
\end{equation}
where $x_i$ are quadratures of the OPOs, $p(t)$ describes the effective laser pumping power (injection minus linear losses), and $C$ corresponds to the strength of saturable nonlinearity. As $p(t)$ grows from a large negative to large positive $p(t) = p_\infty$ values, $E$ anneals from the dominant convex first term on the right-hand side of Eq.~(\ref{E_Ising}) that is minimized at $x_i = 0$ for all $i$, to the minimum of the second term which is the scaled target Ising Hamiltonian with $x_i = \pm \sqrt{p_\infty}$. The temporal change of $p(t)$, therefore, is the annealing parameter combined with gradient descent as
\begin{equation}
{\dot x}_i = - \frac{\partial {E}}{\partial x_i} = C \left( p(t) x_i - x_i^3 \right) + \sum_{i = 1}^{N} J_{ij} x_j. \label{CIM_eq}
\end{equation}
The operation of CIM, therefore, relies on the gradient descent of an annealed energy landscape. All ONN soft spin optimizers exploit this central principle, while the details of the nonlinearity or the gradient dynamics can vary from platform to platform \cite{syed2022bifurcation}. In particular, CIM dynamics is an example of the Hopfield-Tank (HT) network ${\dot x}_i = p(t) x_i + \sum_{i = 1}^{N} J_{ij} x_j$ also used for Ising Hamiltonian minimization \cite{Hopfield1982, Hopfield1985}. Another approach uses second-order resonance to project the XY onto the Ising dynamics \cite{kalinin2018simulating}. In the next section, we describe the principles of gain-based computing and contrast it with simulated and quantum annealing.

\section{Principles of Operation of Gain-Based Computing, Quantum Annealing, and Simulated Annealing}

\begin{figure*}[ht]
    \centering
    \includegraphics[width=2\columnwidth]{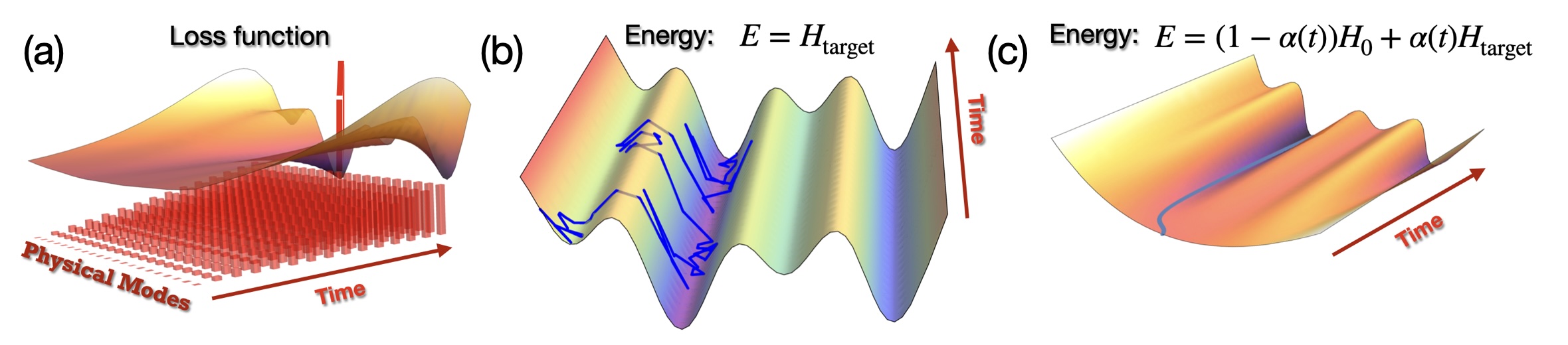}
    \caption{Schematics of the operation of (a) gain-based computing, (b) simulated annealing, and (c) quantum annealing.}
    \label{fig:comparison}
\end{figure*}

Gain-based computing is a computational paradigm where problems are encoded in the gain and loss rates of driven-dissipative systems, as illustrated in Fig.~(\ref{fig:comparison})(a). These systems undergo a symmetry-breaking transition when various physical modes are excited from the vacuum state. As these modes grow, the loss function evolves until a coherent state that minimizes losses emerges. The mode that achieves the minimum of the loss function is amplified, as shown in Fig.~(\ref{fig:comparison})(a). Gain-based computing leverages soft-spin models, which provide enhanced dimensionality and small energy barriers during amplitude bifurcation. Although these models create a complex energy landscape with numerous local minima, making optimization challenging, they are also rich in potential solutions.

Simulated annealing (SA), on the other hand, is a classical optimization technique; see Fig.~(\ref{fig:comparison})(b). SA probabilistically explores the solution space by simulating the cooling of a material to reach a state of minimum energy. It uses thermal fluctuations to escape local minima, as the system trajectory shown in blue indicates, with the probability of accepting worse solutions decreasing over time. This simulates a cooling process that gradually refines the search towards the global minimum. Implemented on classical computing systems using stochastic algorithms, SA explores the energy landscape by thermal fluctuations, with a gradual reduction in temperature controlling the balance between exploration and exploitation. The performance of simulated annealing is influenced by the cooling schedule, which determines how the temperature is reduced over time, as well as the specific parameters of the algorithm.

In contrast, quantum annealing (QA) is a quantum computation method used to find the ground state of a system's energy; see Fig.~(\ref{fig:comparison})(c). QA operates by evolving the system from an initial Hamiltonian, which is usually simple and convex, to the target Hamiltonian that encodes the optimization problem. This evolution relies on the principles of quantum mechanics, specifically quantum tunneling, to explore the energy landscape. Quantum annealing utilizes quantum fluctuations to escape local minima and tunnel through energy barriers, potentially leading to faster convergence to the global minimum. This approach can be advantageous in navigating complex landscapes with high barriers between local minima. In Fig.~(\ref{fig:comparison})(c), the varying energy landscape is shown as the annealing from the initial convex Hamiltonian to the target Hamiltonian takes place in time. The system starts at the ground state of the initial Hamiltonian and remains in the ground state if annealing is sufficiently slow. The blue line shows the system state at each moment.

\section{M\"obius Ladder Graphs}

Cyclic graphs with $N$ nodes are characterized by the circulant coupling matrix $\mathbf{J} \in \mathbb{R}^{N \times N}$, constructed through cyclical permutations of an $N$-vector. These graphs inherently have vertex permutation symmetry, signifying boundary periodicity and uniform neighborhoods. The structure of a circulant matrix is contained in any row, and its eigenvalues and eigenvectors can be analytically derived using the $N$ roots of unity of a polynomial, where the row components of the matrix act as coefficients: $\lambda_{n} = \sum_{j = 1}^{N} J_{1, j} \cos [\frac{2 \pi n}{N} (j - 1)]$ \cite{kalman2001polynomial, zhang2007resistance, gancio2022critical}. We consider the minimization of the Ising Hamiltonian on a particular form of cyclic graph -- M\"obius ladder graphs \nb{with tunable hardness}. These have even $N$ such that the $i$-th vertex has two edges connecting it to vertices $i \pm 1$ with antiferromagnetic coupling with strength $J_{i,i \pm 1} = -1$ (\textit{circle} couplings), and an additional antiferromagnetic coupling with vertex $i + N/2$ with strength $J_{i, i + N/2} = -J$ for $J > 0$ (\textit{cross-circle} couplings). We denote by $S_0$ the state where the spins alternate along the ring so that $s_i s_{i \pm 1} = -1$ for all $i$ (Fig.~(\ref{FigAnalysis})(a)), and by $S_1$ the state where the spins alternate everywhere except at two positions on the opposite sides of the ring: $s_i s_{i + 1} = -1$ for all $i \ne i_0$ and $s_{i_0} s_{i_0 + 1} = s_{i_0 + N/2} s_{i_0 - 1 + N/2} = 1$ (Fig.~(\ref{FigAnalysis})(b)). When $N/2$ is odd, $S_{0}$ is always the ground state with energy $H_{\rm I} (J) = - (J + 2) N / 2$.  When $N/2$ is even, the $S_{0}$ configuration has energy $H_{\rm I}^{(0)}(J) = (J - 2) N / 2$ and $S_1$ has $H_{\rm I}^{(1)}(J) = 4 - (J + 2) N / 2$. Therefore, $S_0$ [$S_1$] is the global minimum (while $S_1$ [$S_0$] is the excited state) if $J < J_{\rm crit} \equiv 4 / N$ [$J > J_{\rm crit}$]. The eigenvalues of the coupling matrix $\bf{J}$ for the M\"obius ladder with $J_{1,j} \in \{-1, 0, -J\}$ are $\lambda_{n} = -2 \cos ( 2 \pi n / N ) - J ( -1 )^{n}$. Equating the two largest eigenvalues $2 \cos(2 \pi / N ) + J$ and $2 - J$ gives the value of $J = J_{\rm e} = 1 - \cos (2 \pi / N)$ at which the leading eigenvectors change. When $J_{\rm e} < J < J_{\rm crit}$ the eigenvalues for $S_{0}$ are less than that for $S_{1}$, despite $S_{0}$ being the lower energy state (see Appendix \ref{Sec: Eigenvectors} for the derivation of the spectra). This is in contrast to computationally simple problem instances, in which the ground state minimizer is located at the hypercube corner of the projected eigenvector corresponding to the largest eigenvalue \cite{kalinin2020complexity}.

\section{Soft-Spin Ising Model}

\begin{figure}[t]
	\centering
	\includegraphics[width=\columnwidth]{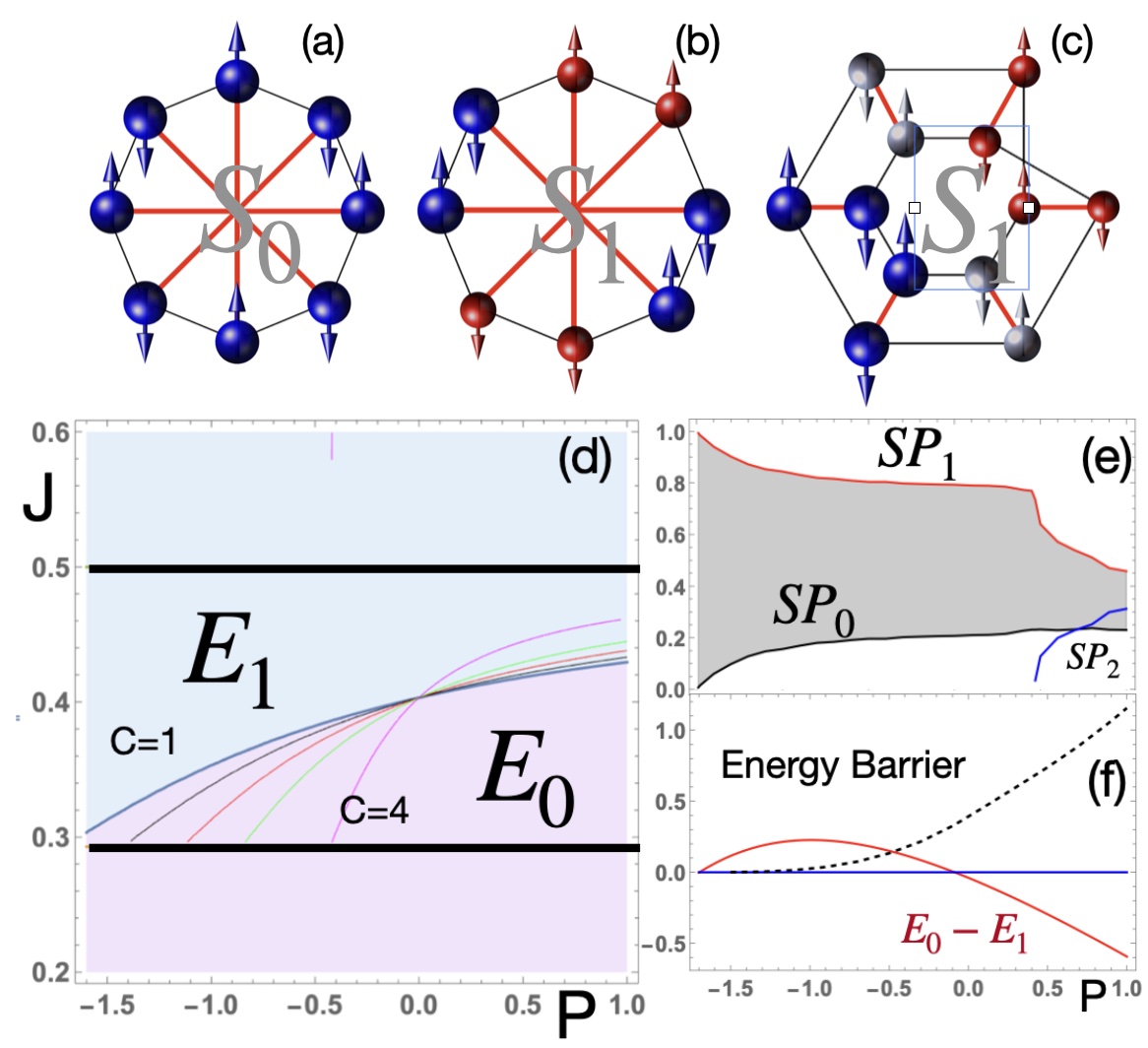}
	\caption{(a-c) Schematic representation of the states realized by soft-spin models of Eq.~(\ref{E_Ising}) on M\"obius ladder graphs with varying cross-circle couplings (shown in red). (a), (b) and (c) depict states that map onto $S_0$, $S_1$ Ising states for $N = 8$ and $S_1$ state for $N = 12$, respectively. (c) uses a different node arrangement that illustrates the graph relationship with the topology of the M\"obius strip. The same colors are used to show equal intensities; the larger sizes correspond to larger intensities. (d) Regions of different global minima of Eq.~(\ref{E_Ising}): $E_1$ in the blue region and $E_0$ in the pink region, in $J - p$ space for $N = 8$ and $C = 1$. Two critical values of $J$ are shown as solid black lines. Between these lines, $S_0$ is expected as the hard-spin Ising model global minimizer. Thin lines show the contours $E_1 = E_0$ for $C = 1$, $1.2$, $1.5$, $2$, and $4$ in that region. (e) Success probability of reaching $E_0$ (labeled as $SP_0$) and $E_1$ (labeled as $SP_1$) states of the soft-spin energy in Eq.~(\ref{E_Ising}) from a point ${\bf x}$ with randomly chosen components $x_i$ in $[-1 , 1]$ for different values of $p$ and $J = 0.4$, $N = 8$, and $C = 1$. For larger values of $p$, a third state of higher energy appears with success probability $SP_2$; when projected on spins $s_i = x_i/|x_i|$ this state corresponds to $S_1$. (f) The height of the minimum energy barrier between $E_1$ and $E_0$ calculated as the energy difference between $E_1$ and the energy of the nearest saddle point is shown as a black dashed line for $J = 0.4$ and $C = 1$. The difference between $E_0$ and $E_1$ is shown in red.}
	\label{FigAnalysis}
\end{figure}

\begin{figure}[t]
\centering
\hspace{-0.6cm}
\includegraphics[width=1.05\columnwidth]{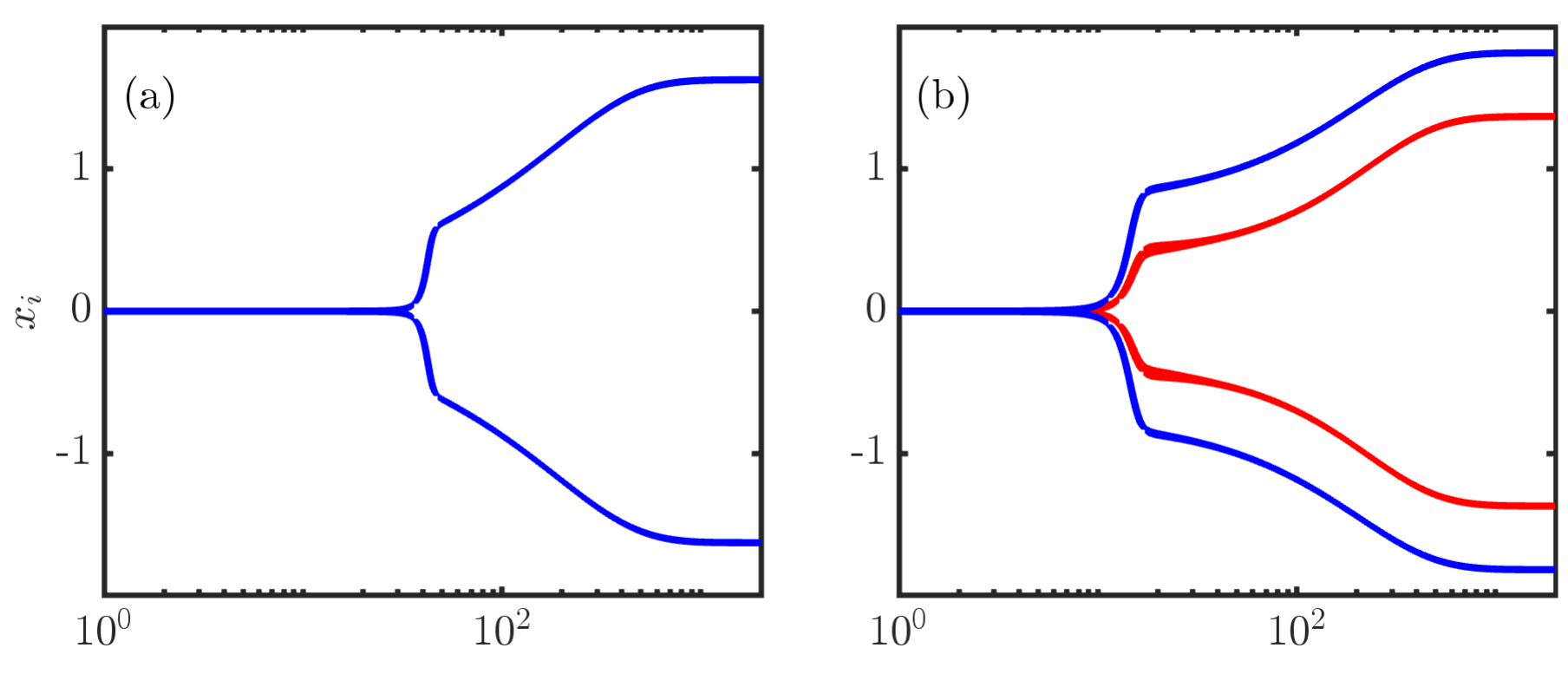}
\caption{(a) Evolution of $N = 8$ soft spins for $J = 0.35$ and (b) $J = 0.55$ according to Eq.~(\ref{CIM_eq}). In each case, the ground state is recovered. The amplitudes connected by the frustrated edges are lower than in the rest of the system and are shown in red. In all runs, $C = 1$, $p_0 = J - 2$, $\varepsilon = 0.003$, $\Delta t = 0.1$, and each $x_i(0)$ is chosen randomly from a uniform distribution in the range $[-0.001, 0.001]$. \label{Fig:CIM-amplitude}}
\end{figure}

\begin{figure}[t]
	\centering
	\includegraphics[width=\columnwidth]{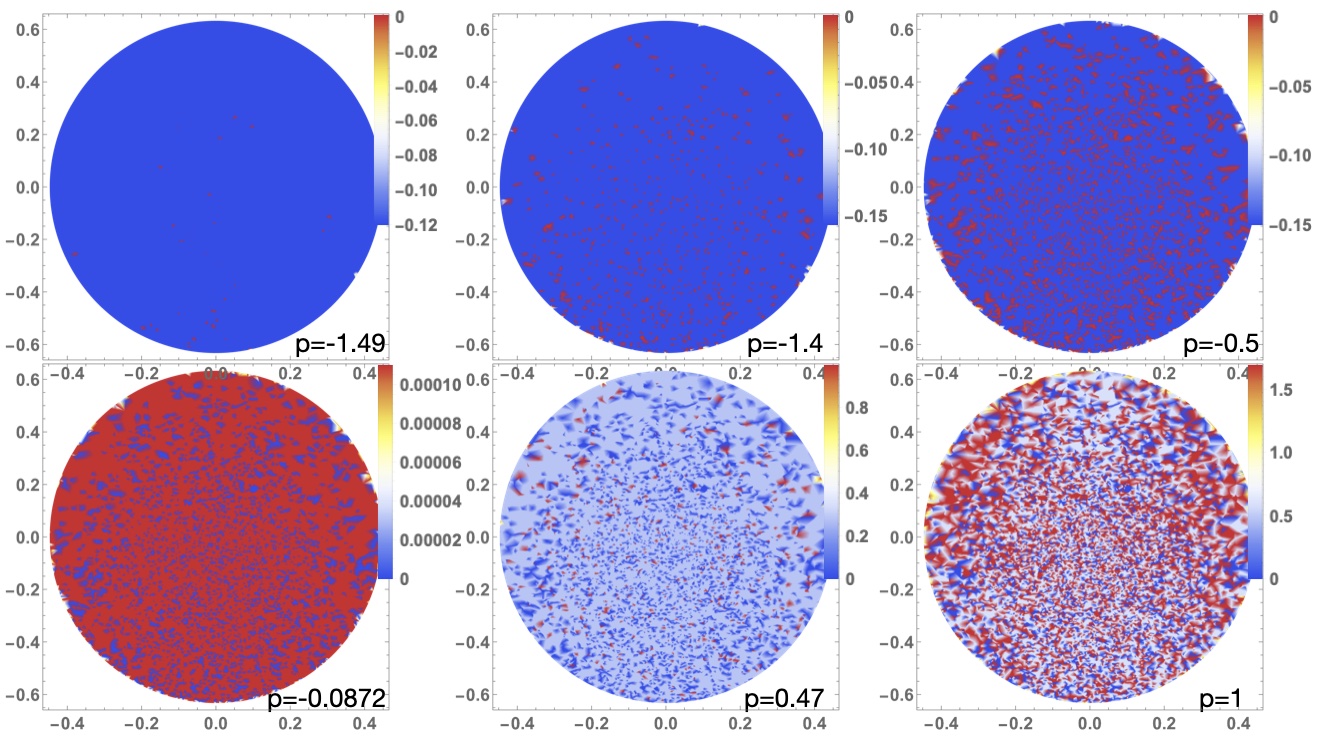}
	\caption{Basins of attraction of the soft spin energy in Eq.~(\ref{E_Ising}) as defined in the main text. We take $N = 8$, $J = 0.4$, $C = 1$, various laser powers $p$, and $20000$ randomly distributed $x_i$ in $[-1,1]$ to show which minimum is reached via gradient descent using Newton's method. To characterize points, the average magnetization $m = \sum_i x_i/N$ (vertical axis) and the correlations along the circle between $x_i$ defined as $X_{\rm corr} = \sum_i (x_i - m)(x_{i+1}-m)/\sum_i (x_i-m)^2$ (horizontal axis), are used. For small $p$, the basin of attraction is dominated by the $S_1$ state as any initial state descends to $E_1$. As $p$ grows, the ratio of the volume of the basins of attraction of the $S_1$  state to the volume of the basins of attraction of the $S_0$ approaches $4$. At the critical value of $p \approx -0.08715$, both $S_1$ and $S_0$ states have the same energy, and after that $S_0$ state becomes the ground state: this is indicated by the switch between ground (blue) and excited (red) states.}
	\label{FigBAa}
\end{figure}

SSIM in Eq.~(\ref{E_Ising}) has real amplitudes $x_i$. As the laser pumping $p(t)$ increases from negative values, the minimizers ${\bf x}^*$ of Eq.~(\ref{E_Ising}) and minima of $E$ change. We associate Ising spins with $x_i$ via $s_i = x_i/|x_i|$. We expect that the soft-spin energy state $E_0$ that corresponds to the  hard-spin Ising state $S_0$, and depicted in Fig.~(\ref{FigAnalysis})(a), is symmetric in amplitudes as all spins experience the same frustration of the cross-circle coupling, so all amplitudes have the same modulus $|x_i| = X.$ From Eq.~(\ref{CIM_eq}), $X$ satisfies $X = \sqrt{p(t) + (2 - J)/C}$, with the corresponding soft-spin energy $E_0 = (J-2)N(2-J+2 C p)/4.$ This state can be realized from a vacuum state when $p(t)$ exceeds $(J-2)/C.$ The soft-spin energy state $E_1$ corresponding to $S_1$, when two side edges are frustrated, is asymmetric in amplitudes. This asymmetry is shown schematically in Figs.~(\ref{FigAnalysis})(b) and (c), in agreement with the dynamical simulations presented in Fig.~(\ref{Fig:CIM-amplitude}). This occurs because the lower energy is achieved if the amplitudes connected by the frustrated edges $|x_i| = X_L$ are lower than in the rest of the system. For $N = 8$ in Fig.~(\ref{FigAnalysis})(b), there are two types of amplitudes: 4 nodes with $\pm |X_L|$ and 4 with amplitudes $|x_i| = X_B$, where $X_B = (1 - J - C p) X_L + C X_L^3$ as obtained from the steady states of equation Eq.~(\ref{CIM_eq}) governing the dynamics of $X_{L}$, while the steady-state on the evolution of $X_L$ gives $(p + 1 + J)X_B + X_L = X_B^3$. By solving the polynomial equation for $X_L$, we can compute $E_1$ across any $p$, $J$, $N$, and $C$. This allows us to discern regions in this parameter space where the global minimum aligns with either $E_0$ or $E_1$ and confirm if these states correspond with the hard-spin Ising Hamiltonian's global minimum. Figure (\ref{FigAnalysis})(d) depicts distinct regions in the $J - p$ parameter space. Within the $J_{\rm e} < J < J_{\rm crit}$ interval $S_0$ emerges as the hard-spin Ising model's lowest energy state. For the soft-spin model, however, only the region shown in pink corresponds to this state ($E_0$). Figure (\ref{FigAnalysis})(d) shows that for values $J_{\rm e} < J < J_{\rm crit}$, as laser power $p$ rises, the $E_0$ state becomes the energy minimum for the soft-spin model, aligning with the hard-spin Ising Hamiltonian's $S_0$. Yet, the success probability of converging to the true ground state does not increase beyond $0.2$ as shown in Fig.~(\ref{FigAnalysis})(e). This is a consequence of increasing amplitudes that bring the increased height of the energy barriers that prevent the system from transitioning to the state $S_0$ (see Fig.~(\ref{FigAnalysis})(f)). Figure~(\ref{FigBAa}) depicts the basins of attraction for various $p$ and fixed $J_{\rm e} < J < J_{\rm crit}.$ The basins of attraction are defined as the sets of points randomly distributed on $[-1, 1]$ from which gradient descent leads to different minima. At the threshold of large negative $p$, the basin of attraction of $E_1$, which is the ground state of $E$ as given by Eq.~(\ref{E_Ising}), dominates. As $p$ increases, the basin of attraction of the excited state $E_0$ increases while at small positive values of $p$, $E_0$ becomes the ground state. With a further increase of $p$, other states with even higher energy appear.

\begin{figure}[t]
	\centering
	\includegraphics[width=\columnwidth]{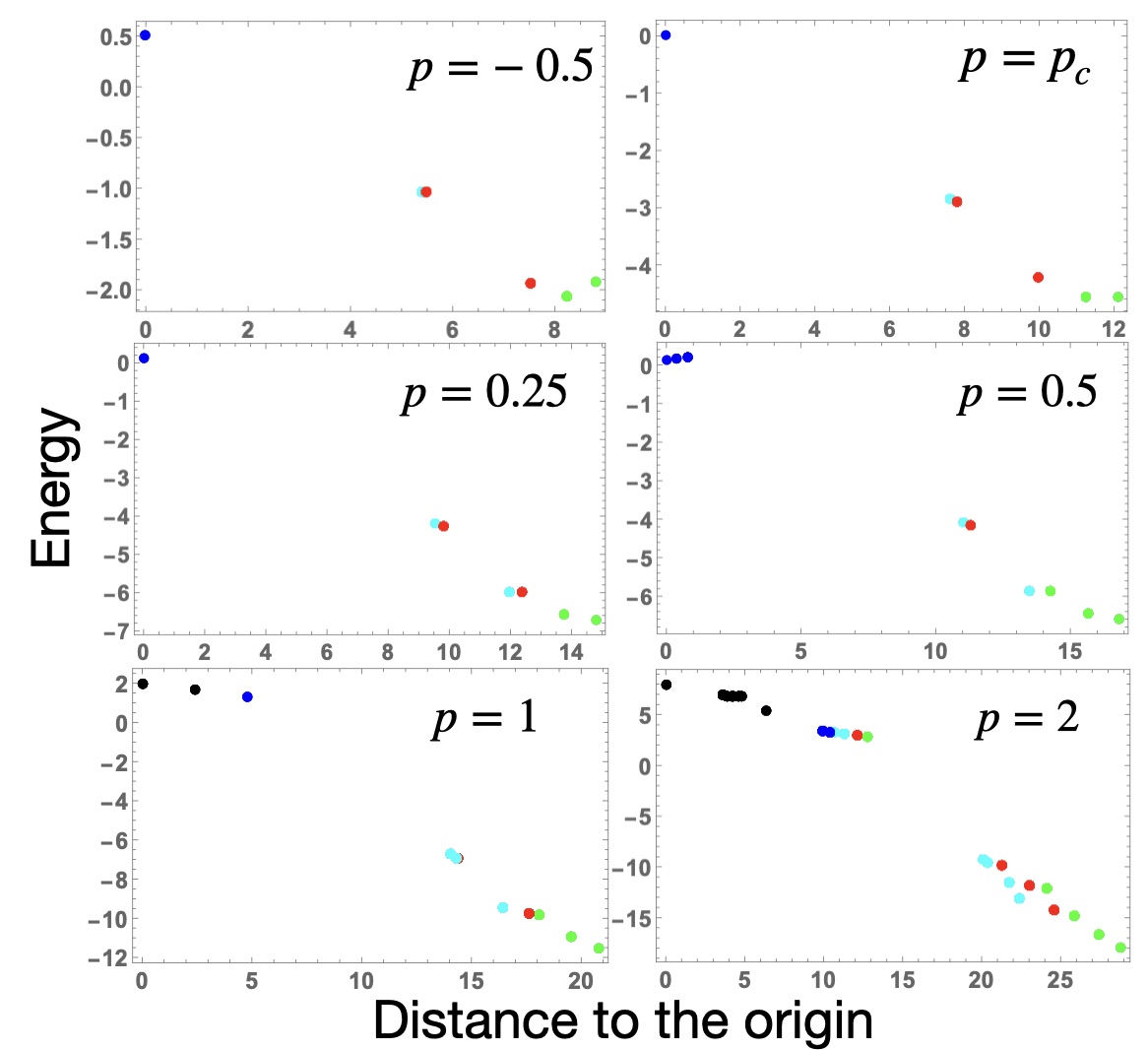}
	\caption{Critical points of the CIM energy (\ref{E_Ising}) for $N = 8$, $C = 1$ and different values of laser power $p$. Minima are shown in green, saddles with one, two, three and 4+ unstable directions are shown as red, light blue, blue and black points, respectively. The $S_0$ state is furthest from the origin, which becomes the global minimum for $p > p_{\rm c} = -0.0872$. The minima for $p = 2$ are $S_0$, $S_1$, $(-,-,+,-,+,-,-,+)$, $(+,-,-,+,+,-,+,-)$ and $(+,-,-,+,+,-,-,+)$ in increasing energy.}
	\label{critical}
\end{figure}

The space structure of soft spin models can be further understood by considering the critical points of their energy landscape for different annealing parameter values \cite{Yamamura2023}. We can determine the critical (minima and the saddle) points by setting ${\partial  E} / \partial x_i = 0$ for all $i = 1, \ldots, N$ and classify them using the Hessian matrix. The number of critical points grows exponentially fast with $p$; however, not in terms of energy and the distance from the  state $x_i = 0$ $\forall$ $i$ as Fig.~(\ref{critical}) illustrates.

The $S_1$ state is always further away from the origin than other critical points ($S_0$ and saddle points). At the same time, the transition between minima $E_0$ and $E_1$ is possible only through a saddle point whose energy relative to $E_1$ and $E_0$ defines the height of the energy barriers; see Fig.~(\ref{FigAnalysis})(f).

\section{Manifold Reduction}

The aforementioned considerations suggest that amplitude heterogeneities have a severely detrimental effect on the optimization process in some regions of parameter space as they allow the soft-spin energy landscape to find and follow its ground state, which is quite different from the ground state of the hard-spin Ising Hamiltonian. This problem has been recognized before \cite{kalinin2018networks, leleu2020chaotic}, but in the context of the final state, so various feedback schemes were suggested to bring all amplitudes to the same value, say $\pm 1$, at the end of the simulations. This could be achieved, for instance, by changing the laser intensity individually for each spin as $\dot{p_i}(t) = \varepsilon (1 - x_i^2)$, where $\varepsilon$ is a small constant parameter. However, as our results on the simple circulant graphs illustrate, this feedback does not change the  most essential part of the dynamics during the pitchfork bifurcation from the vacuum state. Moreover, this feedback becomes important only for amplitudes sufficiently close to $\pm 1$ when the barriers between states are already too high. 

Instead, we suggest introducing feedback restricting the soft spin energy landscape to keep them close to the average value. This restriction can be achieved by modifying the signal intensities bringing them towards the average mass per particle defined by the squared radius of the quadrature $R({\bf x})\equiv \sum_{i=1} x_i^2/N$ as
\begin{equation}
    x_i\rightarrow (1 - \delta) x_i + \delta R x_i/|x_i|.\label{delta}
\end{equation}
If $\delta = 0$, then no adjustment is made. If $\delta = 1$, then all amplitudes are set to the same (average) value. For $0 < \delta < 1$, $1 / \delta$ determines the proportion of the effective space for the restricted evolution.

\begin{figure}[t]
\centering
\includegraphics[width=\columnwidth]{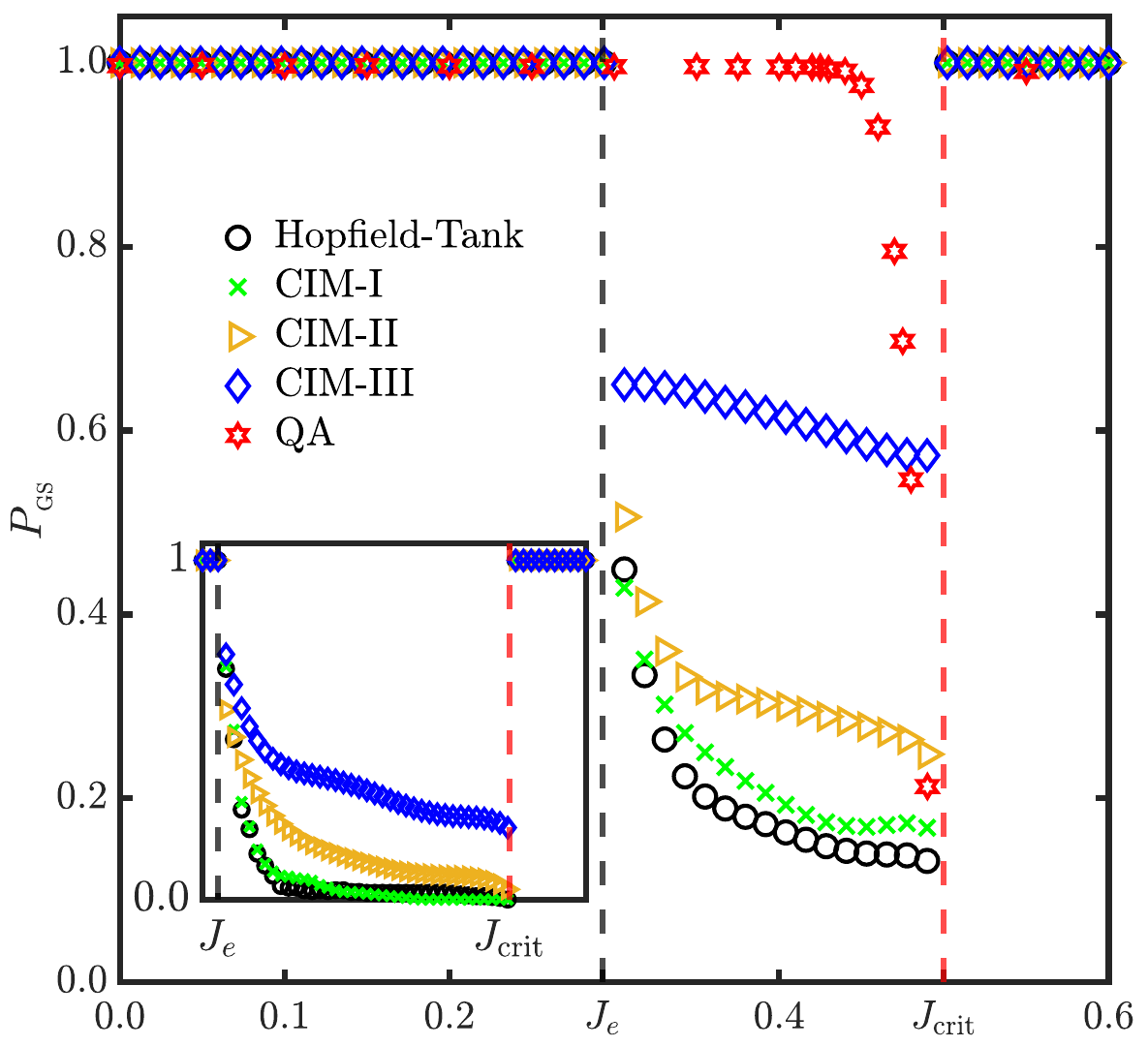}
\caption{Ground state probability for HT, CIM-I, CIM-II, CIM-III, and quantum annealing (QA) for the M\"{o}bius ladder graph with $N = 8$. For CIM-III, for each value of $J$, the optimum value of $0 < \delta < 1$ is chosen based on a set of preliminary runs in which $\delta$ is varied. Two thousand runs are used to calculate the probability of finding the ground state $P_{\rm GS}$ for each value of $J$. For QA, $B = 5$ and $\Delta t = 0.1$. Inset shows the same plots but for $N = 100$. \label{Fig:conv-prob}}
\end{figure}

Figure (\ref{Fig:conv-prob}) shows the probability of finding the ground state of the Ising Hamiltonian using the HT networks: Eq.~(\ref{CIM_eq}) (CIM-I), Eq.~(\ref{CIM_eq}) with individual pumping adjustments $p \rightarrow p_i$ according to $\dot{p_i} = \varepsilon (1 - x_i^2)$ (CIM-II), and Eq.~(\ref{CIM_eq}) with manifold reduction by Eq.~(\ref{delta}) (CIM-III). For CIM-I and CIM-III, we set $p(t) = (1 - p_0) \tanh( \varepsilon \, t) + p_0$. CIM-III shows a significant improvement in finding the ground state compared with other models. Thus, in soft-spin models, the imperative to constrain the manifold implies that dimensional annealing should be tailored according to the energy landscape's characteristics. Quantum annealing, on the other hand, harnesses dimensional annealing within an extended Hilbert space. By utilizing only linear dynamics at the expense of operating within this higher-dimensional phase space, it can effectively navigate energy barriers. Next, we study the quantum evolution on the Ising energy landscape of circulant coupling matrices in order to contrast its performance with soft-spin nonlinear models.

\section{Quantum Annealing}

We consider the transverse field Ising model given by
\begin{align}
\op{H} & = - \frac{1}{2} \sum_{\substack{i,j=1 \\ i \ne j}}^{N,N} J_{ij} \op{S}_i^z \op{S}_j^z - \sum_{i=1}^N h_i \op{S}_i^z - \gamma(t) \sum_{i=1}^{N} \op{S}_i^x \, , \nonumber \\
\op{S}_i^{\alpha} &= \mathds{1} \otimes \mathds{1} \otimes \cdots \mathds{1} \otimes \op{S}^{\alpha} \underbrace{\otimes \mathds{1} \otimes \cdots \otimes \mathds{1} \otimes \mathds{1}}_{i-1 \text{terms}} , \, \, \alpha = x,y,z \, ,
\label{eq:Hamiltonian_QA}
\end{align}
where $\op{S}^{\alpha}$ are the spin-$1/2$ Pauli matrices, $\mathds{1}$ is the $2 \times 2$ identity matrix, and $\otimes$ denotes a tensor product.
The first term, $\op{H}_0$, is diagonal and corresponds to the operator representation of the classical Ising Hamiltonian $H_{\rm I}$; the second term is a symmetry-breaking longitudinal magnetic field; the third term is a transverse field that results in a non-diagonal Hamiltonian operator and gives rise to the quantum Ising model. We will take the annealing term to be of the form $\gamma(t) = B/ \sqrt{t + t_0}$ for some constant $B$ \cite{Kadowaki1998} and set $t_0 = 0.5$.
Our quantum system is made up of $N$ spin-$1/2$ subsystems each having a basis ${\ket{\downarrow}, \ket{\uparrow}}$. A general state $\ket{\Psi}$ of the $N$-spin system can then be written as 
\begin{align}
|\Psi\rangle=\sum_{\xi} C_{\xi}|\xi\rangle \quad \text{with} \quad \sum_{\xi}\left|C_{\xi}\right|^2=1,
\end{align}
where the $C_{\xi}$'s are complex numbers and the basis element 
\begin{align}
|\xi\rangle \equiv \left|\xi_1 \cdots \xi_N\right\rangle=\left|\xi_N\right\rangle \otimes \cdots \otimes\left|\xi_1\right\rangle, \quad \xi_k=\{ \ket{\downarrow},  \,\, \ket{\uparrow} \} \, , 
\end{align}
for $k=1, \cdots, N$. We begin with an initial state, which is the ground state of the transverse field Hamiltonian. The initial state at time $t_i$ can then be expressed as 
\begin{align}
|\Psi(t_i) \rangle = \ket{\psi_{\rightarrow}} \otimes \cdots  \otimes \ket{\psi_{\rightarrow}} \, ,
\end{align}
where for each subsystem $\ket{\psi_{\rightarrow}} = (\ket{\uparrow } + \ket{\downarrow} )/\sqrt{2}$. The wavefunction is then evolved according to the time-dependent Schr\"{o}dinger equation (see Appendix \ref{Sec: Schrodinger} for details) \cite{Norambuena2020}. As $t \rightarrow \infty$,  $\gamma(t) \rightarrow 0$ and the contribution of the last term decays to bring about the target Hamiltonian. Provided $\gamma(t)$ is varied adiabatically, the state evolves whilst remaining in the true ground state of the system, and settles into the target Hamiltonian's desired ground state at sufficiently long times.

\begin{figure}[t]
 \centering
\includegraphics[width=\columnwidth]{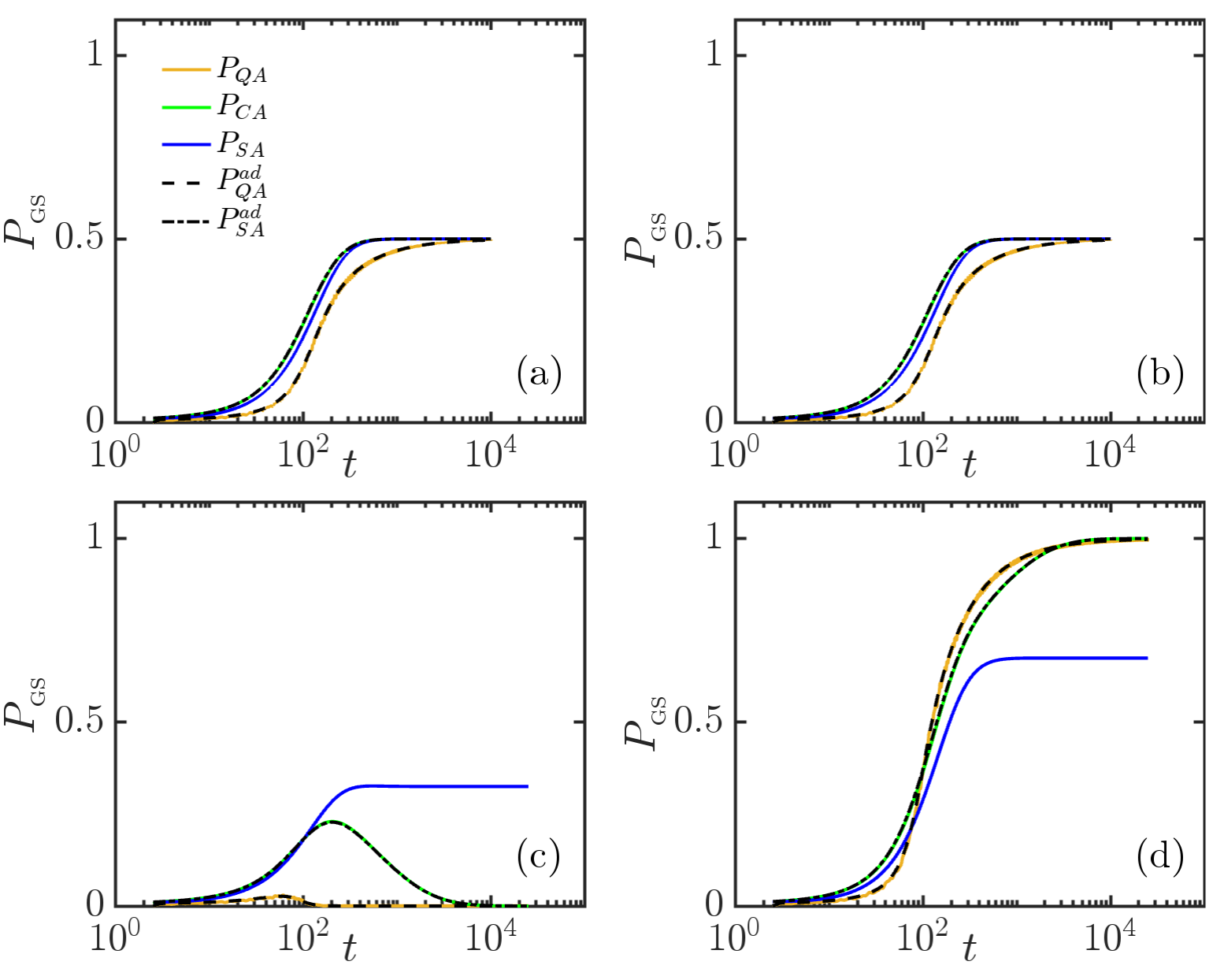}
\caption{Time-evolution of ground state probability of target Hamiltonian with $J = 0$ and $B = 5$ for quantum annealing (QA), single-spin simulated annealing (SA),  classical annealing (CA) and corresponding probabilities expected for adiabatic simulated (SA-ad) and adiabatic quantum (QA-ad) annealing. (a) and (b) correspond to simulation without symmetry breaking term which corresponds to a doubly degenerate ground state. Each figure corresponds to the projection of the probability density onto each one of the ground states; (c) and (d) correspond to a simulation with a symmetry breaking term added which lifts the degeneracy and leads to a unique ground state.
\label{fig:QA_J0}}
\end{figure}

To determine the probability of finding the ground state,  we compute the projection of $\ket{\Psi(t)}$ onto the ground state $\ket{\phi_{_{\rm GS}}}$ of the classical Hamiltonian, $\op{H}_0$, given by $P_{_{\rm GS}}=|\braket{\phi_{_{\rm GS}}}{\Psi(t)}|^2$. In Fig.~(\ref{fig:QA_J0}), we present numerical simulations of the time evolution of the success probability for finding the ground state of an $N = 8$ spin system with $J = 0$ and $B=5$. For comparison, we have also included the results for simulated annealing \cite{Kirkpatrick1983} and classical annealing by evolving a master equation \cite{Kadowaki1998} (see Appendix \ref{Sec: Master} for details). In the former, we allow transition probabilities for single-spin flips only, whereas, in the latter, we allow for all spin flips to reveal the importance of spin correlations on the success probability of finding the ground states. Such collective transitions can be important when topological constraints associated with particular spin configurations can render certain single-spin transitions ineffective at escaping local energy minima.

\begin{figure}[t]
\centering
\includegraphics[width=\columnwidth]{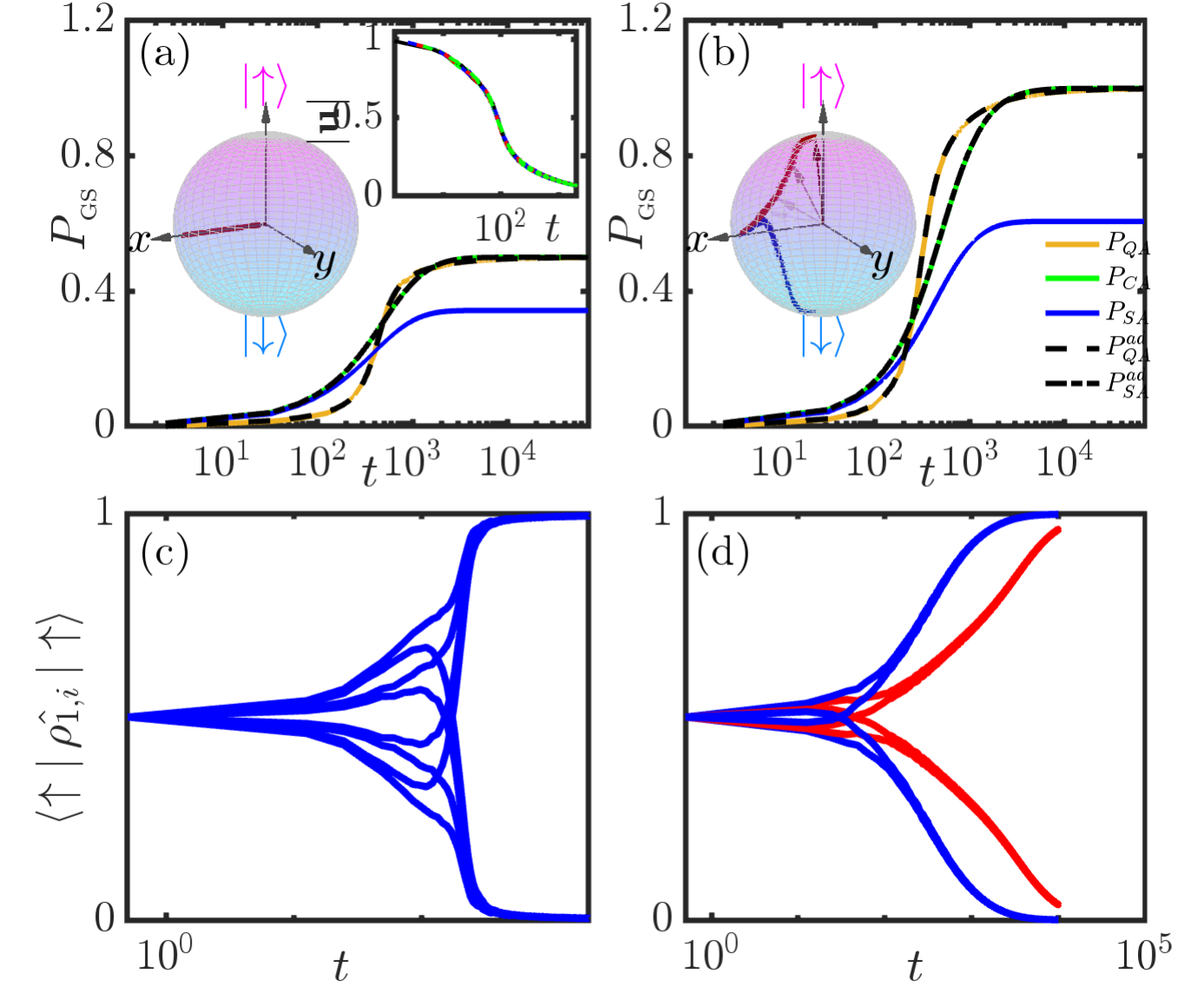}
\caption{Evolution of $N = 8$ spins, with $J = 0.35$, $B = 5$ and $\Delta t = 0.1$;
(a) Ground state probability of target Hamiltonian for quantum annealing (QA), single-spin simulated annealing (SA), classical annealing (CA) and corresponding probabilities expected for adiabatic simulated (SA-ad) and adiabatic quantum (QA-ad) annealing. Insets show Bloch vector for single-spin and magnitude of Bloch vector $|{\mathbf u}|$; (b) similar results as (a) but with symmetry-breaking terms added to Hamiltonian and Bloch sphere shows typical trajectories of two neighboring spins;
; (c) evolution for probability amplitude of $\ket{\uparrow}$ state in quantum annealing simulation with $J = 0.35$ and (d) $J = 0.6$. The amplitudes colored in red correspond to the frustrated spins as in Figs.\ (\ref{FigAnalysis})(b) and (\ref{Fig:CIM-amplitude})(b). \label{Fig:QA_prob_amp}}
\end{figure}


For $J < J_{\rm crit}$, the $S_0$ ground state has a two-fold degeneracy. Therefore, in the absence of a symmetry breaking term, we can expect that the probability of finding one of the ground states is $P_{_{\rm GS}} = 1/2$.
Figures (\ref{fig:QA_J0})(a) and (b) present simulations for the case where no symmetry breaking term is included so that the system contains two degenerate energy minima. Time evolution of the probabilities for finding the system in one of the two degenerate ground states show an equal probability for the system to be found in either one of these states. Moreover, the results are relatively similar regardless of which numerical method is considered. Therefore, quantum annealing, simulated annealing and classical annealing show a similar performance in tracking the ground states as indicated by the curves representing adiabatic evolution of the system.

In Figs.~(\ref{fig:QA_J0})(c) and (d), we present simulations for the case where a symmetry breaking term is included. To introduce the symmetry breaking term, we used $0.05 \ket{\xi}_{S_0} + 0.05 \ket{\xi}_{S_1}$, where $\ket{\xi}_{S_0}$ and $\ket{\xi}_{S_1}$ correspond to the $S_0$ and $S_1$ states, respectively. In contrast to the previous case, the behavior of the different models is now markedly different. In particular, we observe that both quantum annealing and classical annealing correctly evolve with the true ground state as indicated by the curves corresponding to the adiabatic evolution. Moreover, due to the symmetry breaking, there is one unique ground state that the two methods can follow. In contrast, simulated annealing is not always successful at tracking the true ground state even for this case where $J=0$. We found a success probability of only 67\%, whereas the remaining probability is associated with the system converging to what is now a metastable state. These results demonstrate the importance of the symmetry breaking terms and how they affect the time-evolution of the ground state probabilities. 

In Figs.\ (\ref{Fig:QA_prob_amp})(a) and (\ref{Fig:QA_prob_amp})(b), we present results with and without symmetry breaking terms for the case with $J=0.35$ and $B=5$ that corresponds to a hard region of the parameter space for the soft-spin models. As can be seen, now the success probability for simulated annealing degrades even in the absence of symmetry breaking terms. In contrast, classical annealing and quantum annealing continue to perform well. Although the convergence of simulated annealing can be enhanced for slower annealing rates, in general, the success probabilities are lower than the other algorithms we have investigated over a range of annealing schedules (see also Ref.\ \cite{Kadowaki1998}).

To compare the quantum annealing and semi-classical soft-spin simulations, we computed the single-spin reduced density matrix $\op{\rho}_{1,i}$ from the pure state $\ket{\Psi(t)}$. In general, the single-spin density matrix will correspond to entangled states. This is illustrated by recovering the Bloch vector from $\op{\rho}_{1,i}(t)$ (see Appendix \ref{Sec: Bloch} for details). In the inset of Fig.~(\ref{Fig:QA_prob_amp})(a), we show the evolution of the Bloch vector with time evaluated for one of the spins (other spins show similar behavior) for a simulation with $J=0.35$ and $B=5$ in the absence of a symmetry breaking term. We see that the spin is initially aligned along the equator (consistent with the form of $\ket{\rightarrow}$) but shrinks towards the origin as the state evolves. The departure of the Bloch vector from the surface of the Bloch sphere is indicative of quantum entanglement while its dynamics towards the origin signals a spin state that is maximally entangled with the rest of the system. A definite state emerges only upon measurement, which then subsequently collapses the corresponding wavefunction to one specific configuration.

These results demonstrate the striking differences between the states of a fully quantum mechanical description, and a semi-classical description considered earlier. To facilitate comparison with the deterministic semi-classical simulations, we removed the ground state degeneracy in our quantum annealing simulations by introducing the symmetry-breaking term $\op{H}_1$ in Eq.~(\ref{eq:Hamiltonian_QA}). We set $h_i$ to correspond to $0.05 \ket{\xi}_{S_0} + 0.05 \ket{\xi}_{S_1}$. This enforces the evolution towards a specific ground state as can be seen by contrasting the success probability for the ground states presented in Figs.\ (\ref{Fig:QA_prob_amp})(a) and (\ref{Fig:QA_prob_amp})(b). The resulting Bloch vector is shown in the inset of Fig.~(\ref{Fig:QA_prob_amp})(b), and now indicates evolution that ends at the surface of the Bloch sphere, reaching either the $\ket{\uparrow}$ or $\ket{\downarrow}$ state which represents a final state that is not in quantum superposition. Since the Bloch vector does not remain on the surface of the Bloch sphere during the evolution, it clearly demonstrates that though individual spins converge towards a non-correlated value, their evolution bears the imprint of inter-spin correlations. Unlike the semi-classical models, our quantum annealing algorithm consistently identifies the correct ground state for a wide range of parameters in the interval $J_{\rm e} < J < J_{\rm crit}$; (see Fig.~(\ref{Fig:conv-prob})) and demonstrates that correlations play a key role in facilitating the system to converge to the true ground state. However, its performance appears to degrade near $J_{\rm crit}$. \nb{In contrast, the CIM gain-based algorithm is less sensitive near $J_{\rm crit}$ and indicates an advantage of gain-based computing on such a M\"obius ladder graph}. The corresponding time-dependent probability of finding each spin, $i$, in the $\ket{\uparrow}$ state is presented in Fig.~(\ref{Fig:QA_prob_amp})(c) for $J = 0.35$ (and for $J = 0.6$ in Fig.~(\ref{Fig:QA_prob_amp})(d)). As can be seen from the initial evolution of the single-spin probability amplitude, we 
strongly perturb the system towards state $S_1$ through the form of the symmetry breaking terms used. Despite this, the results emphasize the quantum annealing algorithm's capacity to find the correct ground state during gradual $\gamma(t)$ quenches, leveraging the quantum system's expanded phase space.

\begin{figure}[t]
 \centering
\includegraphics[width=\columnwidth]{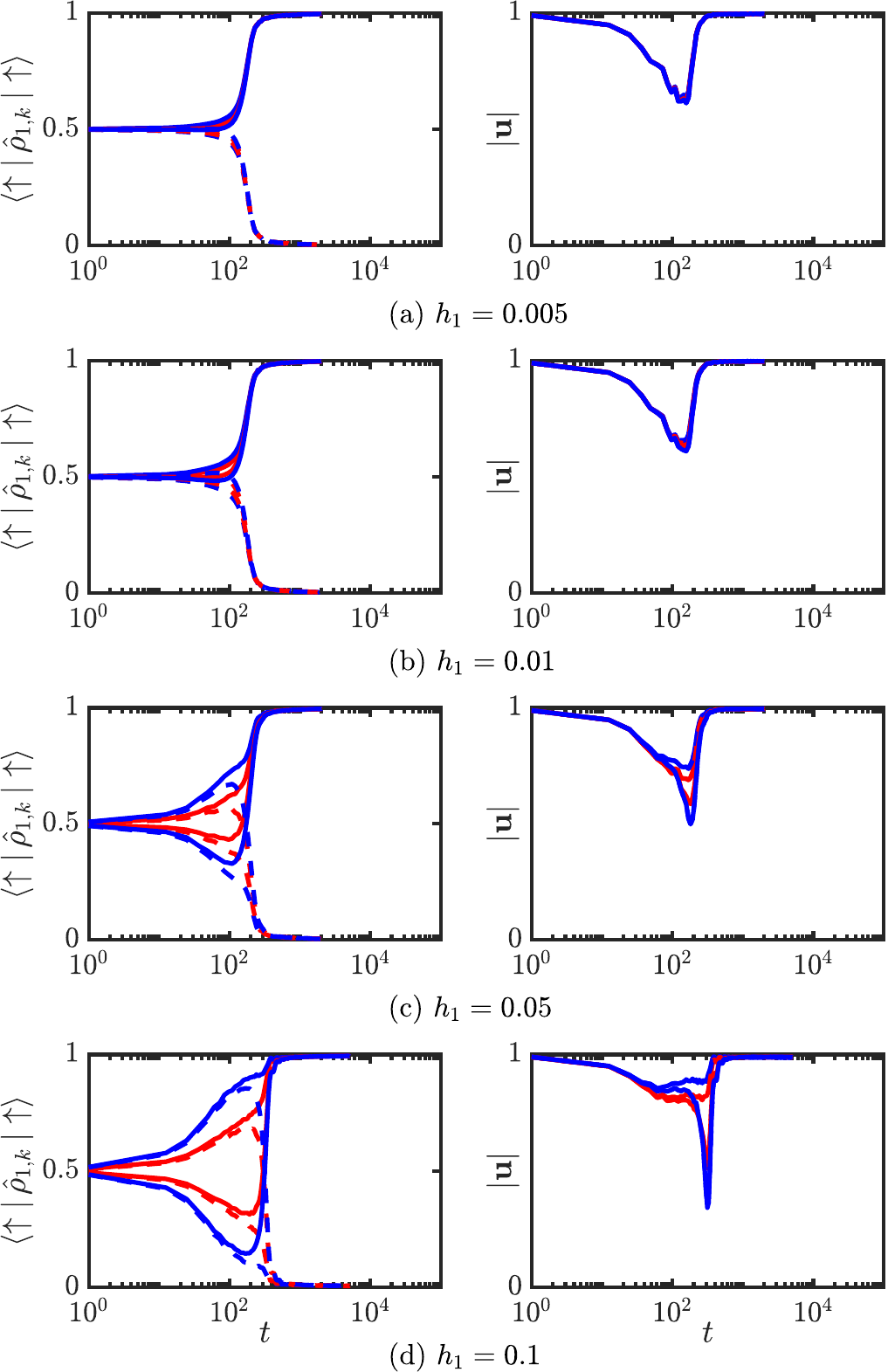}
  \caption{Time-evolution of probability amplitude of $\ket{\uparrow}$ state (left panels) and magnitude of Bloch vectors (right panels) in quantum annealing computation with $J = 0.35$ and $B = 5$. The dashed lines correspond to spins that at the end of the annealing align along the $\ket{\downarrow}$ state whereas solid lines correspond to spins that align with the $\ket{\uparrow}$ state. The red and blue colored lines correspond to the color of the spins shown in Fig.~(\ref{FigAnalysis})(b) and (\ref{Fig:CIM-amplitude})(b). 
  \label{fig:prob_trajectories_J0.35}}
\end{figure}

In order to perform a more systematic study of the impact of including the symmetry breaking terms on the results presented, in Fig.~(\ref{fig:prob_trajectories_J0.35}) we present results for simulations performed by using a symmetry breaking term of the form $0.05 \ket{\xi}_{S_0} + h_1 \ket{\xi}_{S_1}$, where $h_1$ is varied from $0.005$ to $0.1$. For each value of $h_1$, we have evaluated the time evolution of the probability amplitudes $\braket{\uparrow}{\hat{\rho}_{1,k}|\, \uparrow }$ for each spin $k$, as well as the time evolution of the magnitude of the corresponding Bloch vectors $|{\mathbf u}_k|$. The results demonstrate that as $h_1$ is increased, the initial evolution of the probability amplitudes is to align the spins towards the $S_1$ state which is caused by the increasing contribution of the symmetry breaking term. However, as the system navigates the energy landscape, quantum correlations develop as indicated by the decreasing amplitude of the Bloch vectors of the individual spins. This emerging quantum entanglement of the spins prevents the system from becoming stuck in local energy minima and subsequently allows the spins to readjust in order to track the true ground state. Subsequently, the system converges to the true ground state that is well described by a product state as the magnitude of the Bloch vectors converge to unity. We note that during the evolution, the maximal entanglement occurs at the time when the projection of some of the spins flips to the opposite direction. This time also coincides with the time where simulated annealing fails to track the correct ground state in comparison to quantum annealing as reflected in the results of Figs.~(\ref{fig:QA_J0}) and (\ref{Fig:QA_prob_amp}). We, therefore, conclude that quantum correlations play a key role in allowing quantum annealing to outperform other methods in this region of the parameter space of the M\"{o}bius circulant graph.

\section{Conclusions}

In summary, we analyzed the optimization of Ising Hamiltonians, contrasting the classical dynamics of semi-classical soft-spin models with quantum annealing. We discussed the challenges that arise with using semi-classical models, which are due to a broadening dimensionality landscape, especially when the models' global minimum maps to the Ising Hamiltonian's excited state. A solution, termed `manifold reduction', is presented, constraining the soft-spin amplitudes and restricting the dimensionality landscape. On the other hand, we showed that quantum annealing inherently can traverse the Ising Hamiltonian's energy terrain, operating within an extensive Hilbert space. The findings highlight the importance of understanding the influence of dimensionality and the energy landscape overall on optimizing physical systems. Furthermore, they demonstrate how extensions of semi-classical models to include quantum effects has the potential to assist the annealing in navigating the system towards the true ground state.

\section*{Acknowledgements}

J.S.C.~acknowledges the PhD support from the UKRI EPSRC DTP  EP/T517847/1, N.G.B  acknowledges the support from the Julian Schwinger Foundation Grant No. JSF-19-02-0005, HORIZON EIC-2022-PATHFINDERCHALLENGES-01
HEISINGBERG Project 101114978, and Weizmann-UK Make Connection grant 142568.

\appendix

\section{Eigenvectors and Eigenvalues of the M\"obius Ladder Coupling Matrix}
\label{Sec: Eigenvectors}

\begin{figure}[ht]
	\centering
	\includegraphics[width=\columnwidth]{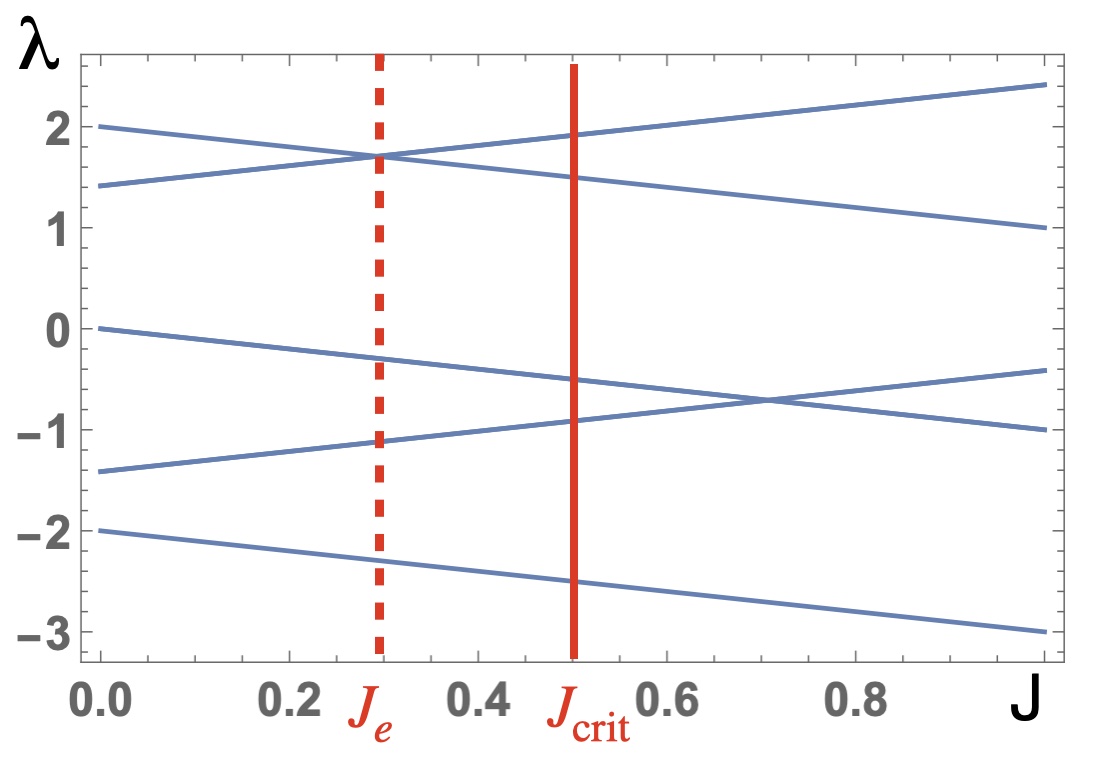}
	\caption{Eigenvalues of an $N = 8$ M\"obius ladder graph as a function of $J$. $J_{\rm e}$ is the value of $J$ where the red dashed line shows the two largest eigenvalues crossing. $J_{\rm crit}$ shows where their energies are equal $E_1 = E_0$. The ground state corresponds to $S_0$ for $J < J_{\rm crit}$ and $S_1$ for $J > J_{\rm crit}$.}
	\label{Je}
\end{figure}

To find the eigenvalues and eigenvectors of the $N \times N$ matrix $\mathbf{J}$ for even $N$, we use the roots of unity, so the solutions of $\omega^N = 1$ are $\omega_k = \exp(i 2\pi k/N)$ for $k=0, \ldots , N - 1$. The  corresponding eigenvectors are $(1, \omega_k, \omega_k^2, \ldots, \omega_k^{N-1})$ which can be  verified by direct computation. Then, from the first row of $\mathbf{J}$, we form the polynomial $f(\omega) = - \omega - J \omega^{N/2} - \omega^{N-1}$ and evaluate it at the unit roots $\omega_k = \exp[i 2 \pi k/N]$ to obtain the eigenvalues
$\lambda_k = f(\omega_k) = - 2 \cos(2\pi k/N) - J(-1)^k$ with the corresponding eigenvectors $v_k = (1, \omega_k, \omega_k^2, \ldots, \omega_k^{N-1}).$ The largest of $f(\omega_k)$ is either $\lambda_{N/2} = 2 - J$ or $\lambda_{N/2 \pm 1} = J + 2 \cos(2\pi/N)$ depending on whether $J < J_{\rm e}$ or $J > J_{\rm e}$ with $J_{\rm e} \equiv 1 - \cos(2\pi/N)$ as the value where these two eigenvalues cross. The corresponding real-valued and mutually orthogonal eigenvectors $\mu_k$ can be formed from $v(\omega_k)$ as $\mu_k = {\rm Re} [v(\omega_k)] + {\rm Im}[v(\omega_k)]$ \cite{zhang2007resistance}. For the two largest  eigenvalues, the eigenvectors are $\mu_{N/2} = (1, -1, 1, -1, \ldots, -1)$ and $\mu_{N/2 \pm 1} = (1,\pm \cos(2 \pi/N)\pm \sin(2 \pi/N), \ldots, \pm \cos(2 \pi k/N)\pm \sin(2 \pi k/N), \ldots, \pm \cos(2 \pi (N-1)/N)\pm \sin(2 \pi (N-1)/N))$. If $N/2$ is even, then $\mu_{N/2\pm 1}$ have the components with two zero values  at the positions separated by $N/2 - 1$ sign alternating components. We illustrate this construction for the M\"obius ladder coupling matrix $\mathbf{J}$ with $N=8$ considered in the main text 
\begin{equation} {\bf J}=
    \left(
\begin{array}{cccccccc}
 0 & -1 & 0 & 0 & -J & 0 & 0 & -1 \\
 -1 & 0 & -1 & 0 & 0 & -J & 0 & 0 \\
 0 & -1 & 0 & -1 & 0 & 0 & -J & 0 \\
 0 & 0 & -1 & 0 & -1 & 0 & 0 & -J \\
 -J & 0 & 0 & -1 & 0 & -1 & 0 & 0 \\
 0 & -J & 0 & 0 & -1 & 0 & -1 & 0 \\
 0 & 0 & -J & 0 & 0 & -1 & 0 & -1 \\
 -1 & 0 & 0 & -J & 0 & 0 & -1 & 0 \\
\end{array}
\right).
\label{JJ}
\end{equation}
The eigenvalues are $\lambda_0=f(\omega_0)=f(1)=-2 -J,$ $ \lambda_1=f(\omega_1)=-\sqrt{2} + J,$ $\lambda_2=f(\omega_2)=-J, $ $\lambda_3=f(\omega_3)=\sqrt{2} + J,$  $ \lambda_4=f(\omega_4)=2- J,$ $\lambda_5=f(\omega_5)=\sqrt{2} + J,$ $ \lambda_6=f(\omega_6)=-J,$ and $\lambda_7=f(\omega_7)=-\sqrt{2} + J$. The eigenvector that corresponds to $\lambda_3$ is $\mu_3=(1,-1,1,-1,1,-1,1,-1)$ and the eigenvector that corresponds to, say, $\lambda_4$ is $\mu_4 = (1,-\sqrt{2},1, 0, -1, \sqrt{2}, -1,0)$. The soft spin system, therefore, follows $(+,-,+,+,-,+,-,-)$ or $(+,-,+,-,-,+,-,+)$ direction at the onset of the pitch-fork bifurcation when $J > J_{\rm e}$ while $\lambda_4$ becomes the dominant eigenvalue of matrix $\mathbf{J}$. Figure~(\ref{Je}) illustrates how these eigenvalues vary as a function of $J$.

\section{Solution of the Time-Dependent Schr\"{o}dinger Equation}
\label{Sec: Schrodinger}

The wavefunction is evolved according to the time-dependent Schr\"{o}dinger equation (with $\hbar=1$) given by
\begin{align} \label{eq:Schrodinger}
i \frac{\dd}{\dd t} \, \ket{\Psi(t)} &= \op{H}(t) \, \ket{\Psi(t)} \, ,\\
\hat{{H}}(t_i) \, |\Psi(t_i)\rangle &= \varepsilon_{g s} \, |\Psi(t_i)\rangle \, ,
\end{align}
where $\varepsilon_{g s}$ denotes the ground state energy of the system at the initial time. To evolve the time-dependent Hamiltonian given by Eq.~(\ref{eq:Schrodinger}), we use a second order accurate Strang time-splitting method where $\op{H}_0$ is evolved for half a time-step $\Delta t$ followed by $\op{H}_1$ for a full time-step and then $\op{H}_0$ for another half a time-step. The resulting time-integration scheme can then be written as
\begin{align}
\ket{\Psi(t_{n+1})} = &\exp \left(-i\frac{\Delta t}{2} \op{H}_D \right) 
\left(-\frac{i}{2} \int_{t_n}^{t_{n+1}} \op{H}_2(\tilde{t}) \dd \tilde{t} \right) \nonumber \\
\times &\exp\left(-i\frac{\Delta t}{2} \op{H}_D \right) \ket{\Psi(t_n)} \, ,  \label{eq:split-step}
\end{align}
where $\op{H}_D = \op{H}_0 + \op{H}_1$ is the diagonal part of the Hamiltonian operator. By placing the Hamiltonian operator $\op{H}_2(t)$ containing the time-dependent term in the middle of the split-step algorithm, we ensure that we have a symmetric time-splitting method. The time integral appearing in Eq.~(\ref{eq:split-step}) was evaluated analytically. In our simulations, we set $\Delta t \equiv t_{n+1} - t_n = 0.1$. The simulations were performed in \texttt{MATLAB}. The exponentials of the diagonal and non-diagonal Hamiltonian can then be readily evaluated using the \verb|expm| function \cite{Norambuena2020}.

\section{Computation of Bloch Vectors in Quantum Annealing Simulations}
\label{Sec: Bloch}

The single-spin reduced density matrix $\op{\rho}_{1,k}$ is obtained  by taking the partial trace of the $2^N \times 2^N$ density matrix $\op{\rho}$ over the Hilbert space of the other $N-1$ spins. For the $k$-th spin, this is defined as
\begin{align}
\op{\rho}_{1,k}(t) = \text{Tr}_{\{N \backslash k \}} \op{\rho} \, ,
\end{align}
where $\{N \backslash k \}$ denotes the $N$ spin Hilbert space excluding the $k$-th spin. The single-spin density matrix can then be parameterized as
\begin{align}
\hat{\rho}_{1,k} 
&= \frac{1}{2} \left( \mathds{1} + \mathbf{u}_k \cdot \hat{\mathbf{S}}\right)
=\frac{1}{2}\left(\begin{array}{cc}
1+w_k & u_k-i v_k \\
u_k+i v_k & 1-w_k
\end{array}\right) \, ,
\end{align}
where ${\mathbf u}_k = (u_k,v_k,w_k)$ defines the corresponding Bloch vector and $\hat{\mathbf{S}} =(\hat{S}^x, \hat{S}^y, \hat{S}^z)$ correspond to the vector of spin-$1/2$ Pauli matrices. For pure states the single-spin reduced density matrix has rank 1, with a magnitude of the Bloch vector $|\mathbf{u}_k|=1$. The surface of the Bloch sphere, therefore, represents all the possible pure states whereas the interior of the sphere corresponds to mixed states. The collapse of the Bloch vector towards the origin implies that the state represents a maximally entangled Bell-like state.

\section{Master Equation for Classical/Simulated Annealing}
\label{Sec: Master}

To model simulated annealing, we follow the method described in Ref.\ \cite{Kadowaki1998}, and introduce the Master equation for the transition probability $P_i(t)$ for each spin configuration as
\begin{align}
\frac{\dd P_i(t)}{\dd t}=\sum_{j=1}^{2^N} A_{i j}(t) P_j(t).
\end{align}
The $2^N \times 2^N$ matrix $A_{i j}(t)$ describes the transition rates. The master equation can be written in the form
\begin{align}
\frac{\dd P_i(t)}{\dd t} &= \sum_{i \ne j} A_{i j}(t) P_j(t) + A_{ii}(t) P_i(t), \\
&= \sum_{i \ne j}  \left( A_{i j}(t) P_j(t) - A_{ji}(t) P_i(t) \right),
\end{align}
where we have made use of the conservation of probability given by
\begin{align}
\frac{\dd }{\dd t} \sum_j P_j(t) = \sum_{i,j} (A_{ji} P_i) = 0,
\end{align}
to arrive at the final equality. Since the normalization condition must hold for any probabilities $P_j$, it follows that
\begin{align}
\sum_{j} A_{ji} = 0 \, , \qquad \text{or} \qquad A_{ii} = - \sum_{i \ne j} A_{ji} \, .
\label{eq_Master_diagonal}
\end{align}
Using Eq.~(\ref{eq_Master_diagonal}) to represent the diagonal terms of the Master equation ensures that a numerical integration of this equation continues to conserve the normalization of the probabilities. The precise form of the transition probabilities is problem specific, although it is common to use the Boltzmann distribution. In our work, we follow Ref.\ \cite{Kadowaki1998} and use the Bose-Einstein distribution such that
\begin{align}
A_{i j}(t)= \begin{cases}\left\{ 
1+\exp \left[\dfrac{\left(E_i-E_j\right)}{ T(t)}\right]\right\}^{-1} \,\, ,& (\text { single-spin flip) } \nonumber \\ -\sum_{k \neq i} A_{k i} \,\, , & (i=j) \\ 0  \, ,& \text {(otherwise) }.
\end{cases}
\end{align}
The form given above that is used for our simulated annealing simulations means that entries of $A_{ij}$ are non-zero only for transitions corresponding to single-spin flips. The annealing is performed by varying the temperature $T(t)$ with time. To maintain consistency with our quantum annealing simulations, we have varied the temperature according to $T(t) = D/\sqrt{t+t_0}$, where $t_0 = 0.5$, and $D$ is a free parameter which we set to $D=5$.

For our classical annealing (CA) simulations, we do not zero out any of the transition probabilities in order to infer how collective transitions of spins at each time-step, as opposed to only single-spin transitions, affects the performance of classical algorithms. It is useful to make the observation that quantum and classical annealing can be closely related to one another if one formulates quantum annealing in imaginary time following a Wick rotation. It then follows that the evolution of the $N$-spin wavefunction $| \Psi(t) \rangle$ is given by
\begin{align}
\frac{\dd}{\dd t}|\Psi(t)\rangle= \left( \mu(t) -\hat{{H}}(t) \right) |\Psi(t)\rangle \, \, .
\end{align}
Here, $\mu(t)$ plays the role of a Lagrange multiplier which ensures that the normalisation of the wavefunction is conserved in analogy with the modification introduced above to the diagonal term of the Master equation. Therefore, by comparing quantum annealing, simulated annealing, and classical annealing, we can distinguish between the effects of retaining all-spin transitions from the difference of evolving our equations in real and imaginary time.


\bibliographystyle{apsrev4-1}
\bibliography{References, ReferencesQA}

\end{document}